# Raman study of lattice vibrations in type II superlattice InAs/InAs$_{1-x}$Sb$_x$

Henan Liu and Yong Zhang[†]

*Optical Science and Engineering Graduate Program and Department of Electrical and Computer Engineering, University of North Carolina at Charlotte, Charlotte, North Carolina, U.S.A*

Elizabeth H. Steenbergen, Shi Liu, Zhiyuan Lin, and Yong-Hang Zhang

*Center for Photonics Innovation and School of Electrical, Computer and Energy Engineering, Arizona State University, Tempe, Arizona, USA*

Jeomoh Kim, Mi-Hee Ji, Theeradetch Detchprohm, and Russell D. Dupuis

*Center for Compound Semiconductors and School of Electrical and Computer Engineering, Georgia Institute of Technology, Atlanta, Georgia, USA*

Jin K. Kim, Samuel D. Hawkins, and John F. Klem

*Sandia National Laboratories, Albuquerque, New Mexico*

[†] Corresponding author: yong.zhang@uncc.edu

**Abstract**

InAs/InAs$_{1-x}$Sb$_x$ superlattice system differ distinctly from two well-studied superlattice systems GaAs/AlAs and InAs/GaSb, in terms of electronic band alignment, common-element at the interface, and phonon spectrum overlapping of the constituents. This fact leads to unique electronic and vibrational properties of the InAs/InAs$_{1-x}$Sb$_x$ system when compared to the other two systems. In this work, we report a polarized Raman study on the vibrational properties of the InAs/InAs$_{1-x}$Sb$_x$ SLs as well as selected InAs$_{1-x}$Sb$_x$ alloys, all grown on GaSb substrates by either MBE or MOCVD, from both growth surface and cleaved edge. In the SL, from the (001) backscattering geometry, an InAs-like LO mode is observed as the primary feature and its intensity is found to increase with increasing Sb composition; from the (110) cleaved edge backscattering geometry, an InAs-like TO mode is observed as the main feature in two cross-polarization configurations, but an additional InAs-like "forbidden" LO mode was observed in two parallel-polarization configurations. InAs$_{1-x}$Sb$_x$ alloys lattice-matched to the substrate (x$_{Sb}$ ~ 0.09) grown by MBE were also found to exhibit the "forbidden" LO mode, implying the existence of some unexpected [001] modulation, but the strained x$_{Sb}$ ~ 0.35 samples grown by MOCVD were found to behavior like a disordered alloy. The primary conclusions are (1) the InAs-like LO or TO mode could be either a confined or quasi-confined mode in the InAs layers of the SL or extended mode of the whole structure, depending on the Sb composition; (2) InAs/InAsSb and InAs/GaSb SLs exhibit significantly different behaviors in the cleaved edge geometry, but qualitatively similar in the (001) geometry; and (3) the appearance of the "forbidden" LO-like mode is a universal signature for SLs and bulk systems resulting from mixing of phonon modes due to structural modulation or symmetry reduction.

**I. Introduction**

Recently, InAs/InAs$_{1-x}$Sb$_x$ type-II superlattices (T2SL) have received considerable attention as a new III-V based IR detection material [1, 2], an alternative to the much more extensively studied InAs/GaSb T2SLs [3-6], that could complement the widely used bulk HgCdTe alloys in middle and long wavelength infrared detection [7-10]. Improved minority carrier lifetimes and dark currents, compared to InAs/GaSb SLs, have been reported for the InAs/InAs$_{1-x}$Sb$_x$ SLs and their devices [11, 12], which is generally believed to be related to the absence of Ga in this system [13, 14]. Therefore, this Ga-free system is becoming a new research interest in the area of IR detection and of basic material physics. In contrast to the considerable recent experimental efforts towards understanding the optical and electrical properties of InAs/InAs$_{1-x}$Sb$_x$ SLs [1, 11, 15-17], very little study is available on their vibrational properties in either experiment or theory, except for some early work on the InAs$_{1-x}$Sb$_x$ alloys [18, 19]. This situation motivates us to conduct a Raman study on the SLs in order to reveal and understand their vibrational properties, and compare with two well studied but distinctly different systems, GaAs/AlAs and InAs/GaSb SLs. This knowledge will fill the gap of our understanding toward InAs/InAs$_{1-x}$Sb$_x$, which represents one of three unique categories of SLs, and lay the foundation for future efforts, such as exploring the mechanism of electron-phonon coupling processes, and using the results in material characterization. Furthermore, this study provides the opportunity to uncover a fundamental and universal effect of structural modulation on longitudinal optical phonons in SLs and other modulated systems.

The InAs/InAs$_{1-x}$Sb$_x$ system is unique, when compared its vibrational properties to those of the two better known SL systems, GaAs/AlAs [20, 21] and InAs/GaSb [22]. The GaAs/AlAs system has type-I or straddling band alignment and the optical phonon spectra of the two constituents do not overlap (similar to the so-called broken-gap type II band alignment) [20], InAs/GaSb has the broken-gap type-II band alignment and the optical phonon spectra of the two constituents overlap with each other but with one enclosing the other (resembling the type I band alignment) [23], whereas InAs/InAs$_{1-x}$Sb$_x$ system has staggered type-II band alignment as well as staggered overlapping optical phonon spectra. Furthermore, InAs/GaSb is special being a no-common-element system, whereas the InAs/InAsSb is a common-cation system. Consequently, the symmetry reduction from D$_{2d}$ in GaAs/AlAs to C$_{2v}$ in InAs/GaSb system does not occur in InAs/InAs$_{1-x}$Sb$_x$ system. Therefore, it is of great interest to investigate InAs/InAs$_{1-x}$Sb$_x$ SLs, which will be beneficial for understanding the SL physics for this particular material system and the field in general.

One of the important SL effects related to symmetry reduction has been that the forbidden LO Raman mode in the (110) back-scattering geometry in bulk becomes allowed, which was first observed in GaAs/AlAs SLs [24, 25], and explained as arising from a standing wave in the SL stacking direction *z* with an effective large wave vector q$_z$ [26]. Recently, the effect was observed in InAs/GaSb by some of us, and speculated as due to the transverse mode of the phonon-polariton [22]. The same effect was also observed in yet another system, a spontaneously ordered GaInP alloy – a monolayer superlattice along the [$\bar{1}$11] direction [27]. There two possible mechanisms were given: the q-dependent Frohlich interaction and the Frohlich interaction due to the electrical field induced by surface charges, in connection with the similar effect reported for a bulk material (GaAs) with doping [28]. We now add two cases, InAs/InAsSb SLs and InAsSb alloys, for this fundamentally a symmetry breaking related phenomenon, which allows us to associate this phenomenon to a universal mechanism – modulation induced mixing of vibration modes.

In this work, we performed polarized micro-Raman measurements on both (001) growth plane and (110) cleaved edge on two sets of InAs/InAs$_{1-x}$Sb$_x$ SL samples grown on GaSb substrates, one by molecular beam epitaxy (MBE) and the other by metal-organic chemical vapor deposition (MOCVD), together with selected InAs$_{1-x}$Sb$_x$ alloy samples, bulk InAs and InSb samples, and alloy and SL samples with Ga doping. Direct comparison of these samples was found to be important to understand the intrinsic vibrational properties of the SLs, and reveal the subtle differences between the SL structures grown by the two growth techniques. Primary findings include (1) an InAs (quasi-)confined LO mode when $x_{Sb}$ is relative low, and its evolution into an extended SL mode when $x_{Sb}$ increases, observed from the front surface measurement; (2) an InAs (quasi-)confined TO mode as well as a "forbidden" LO-like TO mode, observed from the cleaved edge measurement, (3) qualitatively different Raman spectra between InAs/InAs$_{1-x}$Sb$_x$ and InAs/GaSb, when measured from the (110) plane, but qualitatively similar behaviors from the (001) plane; and (4) the appearance of the LO-like mode in the "forbidden" geometry as a common feature in SLs and even bulk alloys when the translational symmetry is broken.

## II. Experiment

All Raman measurements were conducted at room temperature, using a Horiba HR800 confocal Raman microscope equipped with a charge-coupled device (CCD). The Raman signals were collected by a 100× Olympus objective lens with numerical aperture NA = 0.9. The spectrometer was calibrated to yield the Si Raman peak at 520.7cm$^{-1}$. By using a 532nm laser, we were able to achieve a spectral dispersion of 0.44 cm$^{-1}$/pixel, and a spatial resolution of ~ 0.36 µm. To avoid sample heating, a sufficiently low laser power (~ 0.22 mW) was used [22].

Typically, a redshift of ~1.5 cm$^{-1}$ was found for these samples when a factor of 10 times higher laser power was used. Also, two additional Raman modes at ~131 and ~151 cm$^{-1}$ often appeared at the higher power. We do not intend to discuss the details of the higher power results, but wish to point out that these features, reported in the previous study on InAs$_{1-x}$Sb$_x$ alloys [19], were actually from local heating induced formation of Sb elemental crystal [29].

All samples were grown on (001) GaSb substrates, either by MBE or MOCVD, with their structural information listed in Table 1. Because the bulk lattice constant sizes are in the order of $a_{InAs}$ (6.0584 Å) < $a_{GaSb}$ (6.0959 Å) < $a_{InSb}$ (6.4794 Å), InAs$_{1-x}$Sb$_x$ is lattice matched to the GaSb substrate at $x_c$ = 0.089. The SL structure as a whole was targeted to be lattice matched to the substrate, which means that the InAs layer would be under in-plane tensile strain and the InAs$_{1-x}$Sb$_x$ layer under in-plane compressive strain, if they were indeed coherently strained by the substrate. For x > $x_c$, InAs$_{1-x}$Sb$_x$ alloy is expected to be under in-plane compressive strain, if not relaxed. For MOCVD samples labelled as 3-2xxx (grown at Georgia Institute of Technology), the epilayers were grown in a close-coupled showerhead MOCVD reactor system at 100 Torr on GaSb (100) ± 0.04° substrates. The carrier gas used was H$_2$ with the group III precursors being trimethylindium (TMIn) and triethylgallium (TEGa) and the group V precursors being trimethylantimony (TMSb) and arsine (AsH$_3$). The growth was carried out by first growing a 100 nm thick GaSb buffer layer at 580 °C, then the temperature was ramped down to ~460 °C for the growth of SL or InAs or InAs$_{1-x}$Sb$_x$. For the MBE samples labelled as B1xxx (grown at Arizona State University), a GaSb buffer layer of 500 nm was grown at ~ 600 °C, then the temperature was ramped down to grow a 10 nm AlSb layer before growing the InAs$_{1-x}$Sb$_x$ alloy layer or SL. The growth temperatures were in the range of 509 – 522 °C for the alloy samples, and 420 – 464 °C for the SL samples. These samples were all capped with 10 nm AlSb followed by 10 nm

GaSb, except for B1854 that was capped by 100 nm InAs. MBE sample EB3610 (grown at Sandia National Laboratories) was grown at ~ 420 ºC. More growth details can be found elsewhere [30-32]. The compositions of the $InAs_{1-x}Sb_x$ alloys and $InAs_{1-x}Sb_x$ alloy layers in the SLs were determined by X-ray analyses. The individual layers in the SL samples were found being nearly coherently strained by the substrate [31], the two relatively high Sb composition alloy samples were found to be partially relaxed (3-2483, ~75%; 3-2489, ~79%). The details of the X-ray analysis methods can be found in our previous publications [30, 31]. We note that the compositions were derived under the assumption of an abrupt Sb profile. Although the actual Sb profile has been found to be more complex [32], the relative order of the compositions should remain correct among the SL samples if grown by the same technique. Therefore, the general understanding and conclusions on the SLs to be given in this work are not affected by the precise value of the alloy composition and profile.

## III. Results and Discussion

### 1. $InAs_{1-x}Sb_x$ alloys

We have shown in our recent study on InAs/GaSb SLs that the relative Raman cross section of the bulk constituents can offer very helpful hint to the origin of a Raman mode of the SL [22]. We therefore first compare the Raman spectra of InAs and InSb, as shown in **Fig. 1**, with Fig. 1(a) for (001) backscattering and Fig. 1(b) for (110) backscattering. These two geometries yield, respectively, the LO and TO phonon modes at 238.5 and 217.4 $cm^{-1}$ for InAs, and 190.0 and 179.6 $cm^{-1}$ for InSb. For both geometries, the Raman cross-sections of InSb are approximately a factor of 2 larger than those of InAs.

**Fig. 2** shows (001) backscattering Raman spectra for two $InAs_{1-x}Sb_x$ alloy samples: B1784

with $x_{sb}$ = 0.09 in Fig. 2(a), and 3-2489 with $x_{sb}$ = 0.355 in Fig. 2(b), under four polarization configurations: $z(x',x')\bar{z}$, $z(y',y')\bar{z}$, $z(x',y')\bar{z}$, and $z(y',x')\bar{z}$. Here we adopt the conventions that $x'$, $y'$, and $z'$ are defined as $x'$~[110], $y'$~[$\bar{1}$10], and $z'$~[001], with respect to the cubic axes $x$~[100], $y$~[010], and $z$~[001]. If the alloy is viewed as having $T_d$ symmetry in average, the two parallel-polarization configurations are equivalent and both are allowed for the LO-like phonons, and two cross-polarization configurations are forbidden. Indeed, the alloy samples seem to obey the selection rules (Table 2) for $T_d$ symmetry in this scattering geometry. The primary peak in both alloy samples should be the InAs-like LO mode (LO1), although for the higher $x_{Sb}$ samples an InSb-like LO mode (LO2) was also observed, as reported previously for $InAs_{1-x}Sb_x$ alloys [19]. However, B1784 exhibits a small anisotropy in intensity (~20%) between [110] and [$\bar{1}$10], indicating lower symmetry than $T_d$.

**Fig. 3** shows the comparison between the Raman frequencies of the LO1 mode from our alloy samples and the composition dependence of the LO1 mode for fully relaxed alloys given in the previous report [19]: LO1 (cm$^{-1}$) = 238.5 – 32 $x_{Sb}$ (replacing the $x_{Sb}$ = 0 value with the current room temperature value). Apparently, the observed LO1 frequency of ~229 cm$^{-1}$ for the $x_{sb}$ = 0.355 sample lies above the curve for fully relaxed alloys but significantly lower than the predicted curve for alloys under coherency strain with the substrate (the calculation will be described later). The comparison indicates that the epitaxial layers were partially relaxed, which is consistent with the XRD analysis, suggesting that sample 3-2483 and 3-2489 were respectively 75 and 79% relaxed.

The (110) backscattering spectra are shown in **Fig. 4** in four polarization configurations for the same two alloy samples used in Fig. 2. In $T_d$ symmetry, the TO mode is forbidden in (001) backscattering but allowed in (110) backscattering [33]. However, in the previous report, the InAsSb TO modes were actually measured from (001) backscattering, due to the relaxation of the symmetry selection rule in an alloy, but only in a range of $x_{Sb} \leq 0.23$ [19]. Now we are able to examine the TO modes directly in (110) backscattering. For $T_d$ symmetry, the three configurations, $x'(y',z)\bar{x}'$, $x'(z,y')\bar{x}'$, and $x'(y',y')\bar{x}'$ are allowed and equivalent, and $x'(z,z)\bar{x}'$ is forbidden (see Table 2) [33]. We note that despite the three allowed configurations are equivalent in $T_d$, $x'(y',z)\bar{x}'$ and $x'(z,y')\bar{x}'$ are associated with the vibrations in the *x-y* plane, while $x'(y',y')\bar{x}'$ is associated with the vibration along *z* direction. Therefore, any modulation along the *z* direction, for instance, in a SL, will make $x'(y',y')\bar{x}'$ different from the other two. For the alloy sample, the three configurations should remain equivalent in the ideal situation. For the alloy sample B1784 with $x_{sb} = 0.09$, as shown in Fig. 4(a), the $x'(y',z)\bar{x}'$ and $x'(z,y')\bar{x}'$ configurations are indeed very similar as expected. The primary Raman mode at ~215.5 cm$^{-1}$ should be the InAs-like TO mode (TO1) of the InAs$_{1-x}$Sb$_x$ alloy [19]. The weak peak at ~ 226 cm$^{-1}$ is the TO mode of the GaSb substrate, because the epilayer is relatively thin compared to the laser spot size. However, interestingly, the $x'(y',y')\bar{x}'$ configuration in Fig. 4(b) turns out to be very different from the other two, and the $x'(z,z)\bar{x}'$ configuration is also not as weak as one would expect. More peculiar is the appearance of a Raman mode at ~234.5 cm$^{-1}$, close to the LO1 mode that was observed in (001) back-scattering. Thus, although this sample is supposed to be an alloy, it behaves more like a superlattice with some sort of modulation along the [001] direction. We note that the same peak was also observed in the two Ga-doped samples (B1810 and B1814) but weaker relative to TO1. We will come back to discuss the origin of this mode later when presenting the similar phenomenon observed in the SLs.

For the alloy sample 3-2489 with $x_{sb} = 0.355$, as shown in Fig. 4(c) and (d), in contrast to the lower $x_{Sb}$ sample B1784, this sample behaves more like a bulk material of $T_d$ symmetry, and interestingly the LO1 peak is absent. The Raman mode at ~211 cm$^{-1}$ can be assigned as TO1 and the other one at ~ 187 cm$^{-1}$ could be the InSb-like TO mode (TO2) of the alloy, which was not found previously [19]. The TO1 mode frequencies for the alloys are plotted in **Fig. 3(b)**, comparing to the composition dependence of Ref. [19]: TO1(cm$^{-1}$) = 217.4 – 27 $x_{Sb}$ for x ≤ 0.23 (replacing the x = 0 value with the current room temperature value). The contrast between the two samples does not appear to be incidental, because the three lower $x_{Sb}$ samples (see Table 1) behaved qualitatively the same (more discussions will be given later), and two high $x_{Sb}$ samples were also found to be similar. However, we cannot simply attribute the difference to the composition dependence, because one set of samples were grown by MBE, while the other by MOCVD. It is well known that different types of composition modulations may occur in III-V alloys, and the specific form of the modulation depends sensitively on the growth method and condition [34]. Therefore, more systematic investigation is required to understand the microscopic structures of the alloys.

## 2. InAs/InAs$_{1-x}$Sb$_x$ superlattices

**Fig. 5** compares the spectra of the four polarization configurations for two SL samples with very similar structures, 3-2295 ($x_{Sb}$ = 0.230) in Fig. 5(a) and (b), and B1871 ($x_{Sb}$ = 0.239) in Fig. 5(c) and (d), grown respectively by MOCVD and MBE. The former exhibits LO1 = 236. 4 cm$^{-1}$, and the latter LO1 = 235.1 cm$^{-1}$. Their frequencies are substantially higher than that expected for the corresponding free-standing InAs$_{1-x}$Sb$_x$ alloy at the same composition, ~ 230.8 cm$^{-1}$, but matching or close to the expected values for InAs under tensile strain of ~ 236.4 cm$^{-1}$ or InAs$_{1-x}$Sb$_x$ alloy under compressive strain of ~ 234.5 cm$^{-1}$. The strain effects can be calculated using these formula from literature, for instance, Ref. [35]:

$$\delta\omega_{LO} = \delta\omega_H + 2/3\ \delta\omega_S, \qquad\qquad\qquad (1)$$

$$\delta\omega_{TO} = \delta\omega_H - 1/3\ \delta\omega_S, \tag{2}$$

with

$$\delta\omega_H = -\omega_0\ \gamma\ (\varepsilon_{xx} + \varepsilon_{yy} + \varepsilon_{zz}), \tag{3}$$

$$\delta\omega_S = -\omega_0\ \lambda\ (\varepsilon_{xx} - \varepsilon_{zz}), \tag{4}$$

where $\delta\omega_H$ and $\delta\omega_S$ are, respectively, the contributions of hydrostatic and biaxial or uniaxial strain, $\gamma = -(p+2q)/(6\omega_0^2)$ and $\lambda = (p-q)/(2\omega_0^2)$ are the corresponding dimensionless deformation potentials, the strain tensor components are $\varepsilon_{xx} = \varepsilon_{yy} = (a_{GaSb} - a_{InAsSb})/a_{InAsSb}$, $\varepsilon_{zz} = -2C_{12}/C_{11}\ \varepsilon_{xx}$, $C_{12}$ and $C_{11}$ are elastic constants, $\omega_0$ is the corresponding alloy phonon frequency at zero strain. The values for all the parameters were obtained by linearly interpolation between InAs and InSb, except for the composition dependence of the phonon frequency taken from Ref. [19]. The LO1 values for all the SL samples are plotted in Fig. 3(a), in comparison with three calculated curves: the strain-free $InAs_{1-x}Sb_x$ (black), $InAs_{1-x}Sb_x$ (wine) and InAs (green) under the coherency epitaxial strain. Apparently, the LO1 frequencies of SLs mostly fall between the last two curves. Similar to the situation in InAs/GaSb SLs [36, 37], there are three possible mechanisms for the LO1 mode: (1) InAs confined mode, (2) InAs quasi-confined mode, and (3) extended mode. There is yet another possibility, that is, the LO1 is simply the LO mode (mostly) from the strained $InAs_{1-x}Sb_x$ layers, namely an $InAs_{1-x}Sb_x$ (quasi-)confined mode, similar to the case of $Ge_xSi_{1-x}/Si$ SLs grown on Si substrates where the Raman modes were suggested to be originated from the strained alloy layer [38]. By examining the relative intensity of the LO1 mode in two sets of SL samples with respect to bulk InAs, as shown in Fig. 5(e) for G2289, 2295, 2287, and Fig. 5(f) for B1854, 1871, 1775, as well as the LO1 frequencies in Fig. 3(a), we find that with increasing $x_{Sb}$ from $x_{Sb} \sim 0.165$ to $x_{Sb} \sim 0.35$, the LO1 intensity increases significantly (exceeding that in bulk InAs), while the frequency red shifts from that of strain-free InAs (~238.5 cm$^{-1}$) or strained InAs (~236.4 cm$^{-1}$) but remains above that of strained InAs$_{1-x}$Sb$_x$ of the same $x_{Sb}$. The particularly weak signal for B1854 in Fig. 5(f) could be partially due to the presence of the 100 nm InAs capping layer. If the Raman signal were from the InAs$_{1-x}$Sb$_x$ layer, as the first order approximation that the Raman signal is proportional to the fraction of the alloy layer in the superlattice period, the signal intensity would be much below that of the bulk InAs, which is apparently contradicting to the experimental results. These trends suggest that LO1 is likely an InAs confined or quasi-confined mode in the InAs layers when $x_{Sb}$ is relatively low (e.g., near 0.16), and becomes an extended SL mode for the higher $x_{Sb}$ values. In the higher $x_{Sb}$ region, the InAs confined or quasi-confined mode might continue to exist, but could not be resolved due to the smaller Raman cross section of InAs and perhaps other effects such as structural imperfection. The situation is similar to the case of InAs/GaSb where the theoretically predicted InAs confined or quasi-confined modes are not observable [22]. It is worth mentioning that in one Ga-doped sample (B1818) with $x_{Sb}$ = 0.35, an additional weaker Raman mode at ~ 240 cm$^{-1}$, close to that of InAs (~238.5 cm$^{-1}$), was observed on the higher frequency side of the main peak, as shown in the inset of Fig. 5(f). Confined or quasi-confined modes in the InAs$_{1-x}$Sb$_x$ layers are unlikely to occur unless in a very high $x_{Sb}$ value [36, 37], which means that the observed superlattice LO1 mode cannot be the pure InAs$_{1-x}$Sb$_x$ alloy mode that would be given by the strained InAs$_{1-x}$Sb$_x$ curve in Fig. 3(a). This understanding is expected to be relevant to the somewhat similar system Ge$_x$Si$_{1-x}$/Si SL [39].

    The polarization dependence of the (001) scattering shown in Fig. 5 indicates that the InAs/InAs$_{1-x}$Sb$_x$ SL obeys the Raman selection rules (see Table 2) for the SL with D$_{2d}$ symmetry with these four allowed Raman tensors, given in the basis of ($x,y,z$) as A$_1$ = [(a,0,0), (0,a,0), (0,0,b)], B$_2$($z$) = [(0,d,0), (d,0,0), (0,0,0)], E($x$) = [(0,0,0), (0,0,e), (0,e,0)], E($y$) = [(0,0,e), (0,0,0), (e,0,0)] [24, 40]. In general, the (001) back-scattering spectra of the InAs/InAs$_{1-x}$Sb$_x$ SLs are

rather similar to those of InAs/GaSb [22]. Namely, they primarily show one major peak with broadening toward the lower frequency side, obeying the similar Raman selection rules as in the bulk. Therefore, they do not offer as much information about the vibrational properties of the SL as the spectra from the (110) cleaved edge [22].

We now examine the cleaved edge results. **Fig. 6** shows the (110) back-scattering spectra of three SL samples in four polarization configurations for the same two samples ($x_{Sb}$ = 0.230, 0.239) used in Fig. 5(a) and (b), and another one with a higher composition $x_{Sb}$ = 0.33 (EB3610) and also from a difference source. The polarization dependence is found to be qualitatively similar to the MBE alloy sample with x = 0.09 shown in Fig. 4(a). The two cross-polarization configurations, $x'(y',z)\bar{x}'$ and $x'(z,y')\bar{x}'$, are related to the TO modes that vibrate in the ($x,y$) plane, described by the Raman tensor E($x$) or E($y$). The primary Raman peak is InAs-like TO mode, referred to as TO1, and the much weaker peak is InSb-like TO mode (TO2). The TO1 mode frequencies for all SLs are summarized in Fig. 3(b), compared with the values for the strain-free $InAs_{1-x}Sb_x$, strained $InAs_{1-x}Sb_x$ and strained InAs. The intensities of the TO1 mode were found to be comparable to but weaker than that of the bulk InAs. However, the intensity comparison is less reliable for the cleaved edge measurement, because of the overall small thickness of the SL region. The frequencies of the TO1 modes are substantially higher than the corresponding values of the strained $InAs_{1-x}Sb_x$ alloys. Thus, this mode is unlikely originated from the strained alloy layers. For the relatively low $x_{sb}$ sample (e.g., x ~ 0.16), TO1 frequency is very close to that of strained InAs, thus, TO1 is likely confined or quasi-confined TO mode in the InAs layers. For the higher $x_{Sb}$ samples, the TO1 frequencies fall between the two limits, as shown in Fig. 3(b), thus, it is reasonable to assume they are extended modes of the SL as a whole. These assignments are similar to the LO1 mode obtained from the (001) backscattering.

In the two parallel configurations, $x'(y', y')\bar{x}'$ and $x'(z, z)\bar{x}'$, a peak close to the LO1 frequency is observed. For the (001) SL, both $A_1$ and $B_2(z)$ modes, with vibration along $z$, are allowed for $x'(y', y')\bar{x}'$ and only $A_1$ mode is allowed for $x'(z, z)\bar{x}'$ (see Table 2) [24]. $B_2(z)$ is originated from the deformation potential, and $A_1$ from the Frohlish interaction, which qualitatively explains the intensity difference between the two configurations [24]. In fact, including the earlier observations in GaAs/AlAs [24] and spontaneously ordered GaInP alloy [27], and the recent finding in InAs/GaSb SLs [22], the appearance of a LO mode in the "forbidden" geometry has now been established as a common feature for the SL structures. A simple explanation of LO1-like mode in this geometry can be given as that the phonon mode with a small wave vector along $x'$ (to satisfy the momentum conservation) has a standing wave motion along the $z$ direction (with $q_z = 0$) due to zone folding in the SL, thus, the participating phonon mode has a large effective wave vector in the $z$ direction [26]. Technically this is a TO mode with a small SL wave vector along $x'$ to satisfy the momentum conservation but with large $q_z$ components of the bulk phonon modes, whereas the (001) scattering involves a LO mode with a small SL wave vector along $z$ but nevertheless may have large $q_z$ components of the folded bulk phonon modes. Therefore, the LO modes observed in the two scattering configurations could involve the similar components of the folded bulk phonon modes but with different SL wave vectors, respectively, along the z and $x'$ direction, and thus their frequencies are slightly different and could be viewed as the longitudinal and transverse modes of the phonon-polariton in the SL.

**Fig. 7** compares the spectra of the $x'(y', y')\bar{x}'$ configuration, normalized to LO1, for two set of samples, G2289, 2295, 2287 and B1854, 1871, 1775, respectively in Fig. 7(a) and (b), with their peak intensity ratios between LO1 and TO1 plotted in Fig. 7(c). Interestingly, we find that the intensity ratio of the LO1 mode to the TO1 mode increases with increasing the Sb composition of the $InAs_{1-x}Sb_x$ layer, which could be understood as due to enhanced modulation in elastic and electronic properties with increasing contrast between the InAs and $InAs_{1-x}Sb_x$ layer. It appears that this intensity ratio can serve as an empirical measure of the deviation from the bulk as a result of vertical structural modulation. With this understanding, we may speculate that the results observed for the alloy sample with $x_{Sb} = 0.09$ suggest the possible existence of unintended vertical modulation. Although such modulation might not be periodic as in a SL (thus no standing wave formation), the perturbation seems to be sufficient to induce some phonon scattering effects that may also lead to the mixture of the modes with different $q_z$ values. Spontaneous composition modulation along the $z$ axis has been reported in $InAs_{1-x}Sb_x$ alloys with $0.4 \leq x \leq 0.8$ [41], although not in such a low composition. More careful structural study is required to identify the exact nature of the modulation, but cleaved edge polarized Raman has been shown to be a very sensitive tool for revealing the existence of the modulation.

**IV. Comparison between InAs/$InAs_{1-x}Sb_x$, InAs/GaSb, and GaAs/AlAs superlattices**

In the (001) backscattering geometry, the selection rules in the SLs for the four commonly adopted configurations $z(x',x')\bar{z}$, $z(y',y')\bar{z}$, $z(x',y')\bar{z}$, and $z(y',x')\bar{z}$ are usually the same as in the bulk. It is rather unique that for GaAs/AlAs SLs multiple confined LO modes can be observed benefiting from their LO phonon spectra being well separated [42]. For both InAs/GaSb SLs [22] and InAs/$InAs_{1-x}Sb_x$ SLs (Fig. 5), there is only one primary LO mode. Therefore, the Raman scattering results of this geometry are not very informative for understanding the vibrational properties of the SLs, and no qualitatively difference is revealed between the two systems.

It is the (110) backscattering geometry that has revealed some interesting and subtle differences between these two systems. To show clearly the similarity and difference, in Fig. 8 we compare the typical spectra of the two systems (using sample B1871 and of sample IFA in Ref. [22]) measured under comparable conditions in four important configurations. For $z(x', x')\bar{z}$ of the (001) backscattering, Fig. 8(a) shows only one primary LO mode for each SL. However, there are some interesting differences: for the InAs/InAs$_{1-x}$Sb$_x$ SL, an extended InAs-like LO mode (LO1) that is stronger than that of bulk InAs (Fig. 5(f)); for the InAs/GaSb SL, a quasi-confined GaSb-like LO mode that is weaker than that of the bulk GaSb, with a further weaker GaSb-like TO mode, but no expected confined InAs LO mode [22]. For the (110) backscattering, the two cross-polarization configurations $x'(y', z)\bar{x}'$ and $x'(z, y')\bar{x}'$ are always very similar for all above mentioned SLs ([22, 24, 25] and Fig. 6), as dictated by the symmetry and associated with the TO mode with vibration along y'. Thus, Fig. 8(b) – (d) compare the three configurations of the (110) backscattering between the two systems: $x'(y', z)\bar{x}'$, $x'(y', y')\bar{x}'$, and $x'(z, z)\bar{x}'$. For $x'(y', z)\bar{x}'$ of Fig. 8(b), the InAs/GaSb SL exhibits a GaSb confined TO mode, an InAs-like quasi-confined TO mode, and a weak InSb-like interface mode; whereas the InAs/InAs$_{1-x}$Sb$_x$ SL shows an extended InAs-like TO1 mode and a weak, confined InSb-like TO2 mode. For $x'(y', y')\bar{x}'$ of Fig. 8(c), the InAs/GaSb SL exhibits a rather different spectrum from the cross-polarization, namely with primarily an extended TO mode; whereas the InAs/InAs$_{1-x}$Sb$_x$ SL yields the same TO features as in the cross-polarization, similar to the case of GaAs/AlAs SLs [24, 25]. And for all the three SL systems, a "forbidden" LO-like mode appears in this configuration. Furthermore, for $x'(z, z)\bar{x}'$ of Fig. 8(d), which is a forbidden configuration under T$_d$ symmetry but allowed for the SL A$_1$ mode, the LO-like mode as well as other modes are usually very weak compared to $x'(y', y')\bar{x}'$, as in the cases for the InAs/GaSb SL and GaAs/AlAs SLs [22, 24]. However, for the InAs/InAs$_{1-x}$Sb$_x$ SL the intensity of the LO-like mode is about one half of that in $x'(y', y')\bar{x}'$. It turns out that the situation in InAs/InAs$_{1-x}$Sb$_x$ SLs is actually very similar to that of spontaneously ordered GaInP alloy on the cleaved edge [27].

## V. Summary and Conclusions

In summary, we have observed several intrinsic vibrational features in InAs/InAs$_{1-x}$Sb$_x$ SLs. In the (001) backscattering geometry, a confined or quasi-confined InAs LO mode (LO1 mode) has been revealed when x$_{sb}$ is near 0.16, which then evolves into an extended SL mode for higher x$_{sb}$ values. In the (110) backscattering geometry, two modes have been revealed: an InAs-like TO mode (TO1) with E(*x*) or E(*y*) symmetry that also evolves from an InAs confined or quasi-confined mode into an extended mode with increasing x$_{sb}$, and a LO1-like mode with A$_1$ and B$_2$(*z*) symmetry. This LO-like mode has now been established as a common feature that can be observed in the back-scattering geometry from the plane containing the axis of the structural modulation, for instance, the (110) plane for a [001] superlattice as in GaAs/AlAs, InAs/GaSb, and InAs/InAs$_{1-x}$Sb$_x$ or the (110) plane for a CuPt ordered GaInP along [$\bar{1}$11]. A unified understanding is given for all these seemingly very different types of SLs and alloys as resulting from phonon mode mixing associated with either a structural or certain form of modulation that breaks the translational symmetry. Besides, InAs/InAs$_{1-x}$Sb$_x$ and InAs/GaSb SLs were shown to exhibit qualitatively different spectroscopy signatures when probed from the (110) cleaved edge, but not from the (001) growth plane. The InAs$_{1-x}$Sb$_x$ alloys grown by MBE with x$_{Sb}$ lattice-matching to the GaSb substrate were found to possibly have some structural modulation along the growth direction. Additionally, in InAs$_{1-x}$Sb$_x$ alloys, Ga-doping effects were also briefly examined, and two previously reported unexplained peaks were found to be the result of unintended laser induced formation of Sb elemental crystal.

**Acknowledgement**

The work at UNCC, GT, and ASU was supported by ARO/MURI (W911NF-10-1-0524, Dr. William Clark). Sandia National Laboratories is a multi-mission laboratory managed and operated by Sandia Corporation, a wholly owned subsidiary of Lockheed Martin Corporation, for the U.S. Department of Energy's National Nuclear Security Administration under contract DE-AC04-94AL85000. YZ would like to thank Dr. Andrew Norman of NREL for very helpful discussions, and acknowledge the support of Bissell Distinguished Professorship at UNCC.

**Table 1**. Sample information

| Sample # | InAs/InAs$_{1-x}$Sb$_x$ (nm) | Sb composition $x$ | Total thickness (µm) | Note |
|---|---|---|---|---|
| B1854 | 7.7/2.4 | 0.205 | 2.5 | 100 nm InAs cap |
| B1871 | 5.2/4.7 | 0.239 | 0.9 | |
| B1775 | 15.3/4.7 | 0.351 | 0.96 | |
| B1816 | 15.3/4.7 | 0.35 | 0.96 | InAsSb:Ga (center) |
| B1818 | 15.3/4.7 | 0.35 | 0.96 | InAsSb:Ga (top and bottom) |
| 3-2287 | 13.3/5.59 | 0.255 | 0.57 | |
| 3-2289 | 7.04/2.27 | 0.165 | 0.28 | |
| 3-2295 | 5.29/4.80 | 0.230 | 1 | |
| EB3610 | 4.6/1.7 | 0.33 | 0.63 | |
| B1784 | InAsSb | 0.09 | 1 | |
| B1810 | InAsSb | 0.09 | 1 | InAsSb:Ga (top and bottom of the SL region) |
| B1814 | InAsSb | 0.09 | 1 | InAsSb:Ga (middle of InAsSb layers) |
| 3-2483 | InAsSb | 0.343 | 0.3 | 75% relaxed |
| 3-2489 | InAsSb | 0.355 | 0.5 | 79% relaxed |
| 3-2296 | InAs | | | |
| wafer | InSb | | | |

**Table 2**. Raman selection rules for $D_{2d}$ and $T_d$ Raman modes in (001) and (110) back-scattering geometries. The notations *x*, *y*, *x′*, *y′*, and *z* refers to *x*~[100], *y*~[010], *x′*~[110], *y′*~[$\bar{1}$10], and *z*~[001], respectively.

| Symmetry Group | $D_{2d}$ | | | | $T_d$ | | |
|---|---|---|---|---|---|---|---|
| Symmetry / Geometry | E(*x*) | E(*y*) | $B_2$(*z*) | A1 | $F_2$(*x*) | $F_2$(*y*) | $F_2$(*z*) |
| $z(x', x')\bar{z}$ | 0 | 0 | $d^2_{LO}$ | $a^2_{LO}$ | 0 | 0 | $d^2_{LO}$ |
| $z(y', y')\bar{z}$ | 0 | 0 | $d^2_{LO}$ | $a^2_{LO}$ | 0 | 0 | $d^2_{LO}$ |
| $z(x', y')\bar{z}$ | 0 | 0 | 0 | 0 | 0 | 0 | 0 |
| $z(y', x')\bar{z}$ | 0 | 0 | 0 | 0 | 0 | 0 | 0 |
| $x'(y', y')\bar{x}'$ | 0 | 0 | $d^2_{TO}$ | $a^2_{TO}$ | 0 | 0 | $d^2_{TO}$ |
| $x'(z, z)\bar{x}'$ | 0 | 0 | 0 | $b^2_{TO}$ | 0 | 0 | 0 |
| $x'(y', z)\bar{x}'$ | $e^2_{TO}/2$ | $e^2_{TO}/2$ | 0 | 0 | $d^2_{TO}/2$ | $d^2_{TO}/2$ | 0 |
| $x'(z, y')\bar{x}'$ | $e^2_{TO}/2$ | $e^2_{TO}/2$ | 0 | 0 | $d^2_{TO}/2$ | $d^2_{TO}/2$ | 0 |

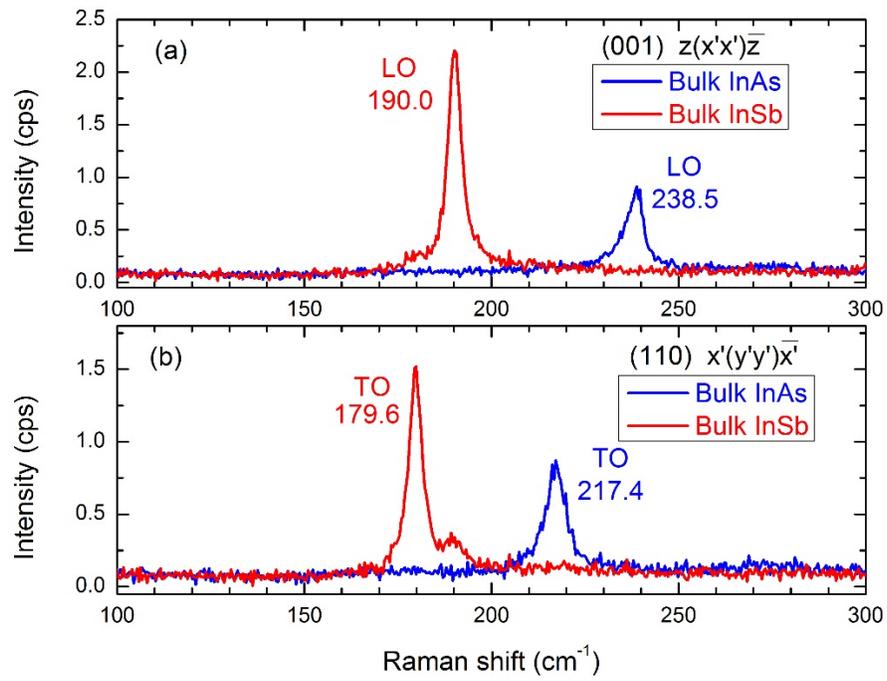

**Fig. 1**. Raman spectra of bulk InAs and InSb. (a) (001) backscattering, (b) (110) backscattering.

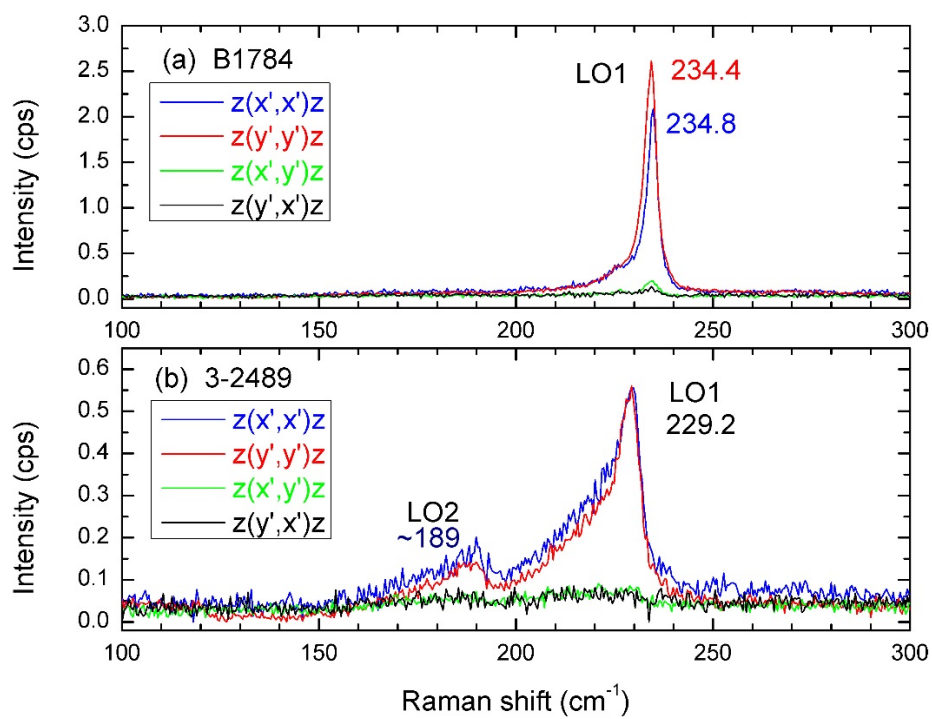

**Fig. 2**. (001) backscattering Raman spectra of InAs$_{1-x}$Sb$_x$ alloys in four polarization configurations. (a) B1784 (x$_{Sb}$ = 0.09) grown by MBE. (b) 3-2489 (x$_{Sb}$ = 0.355) grown by MOCVD.

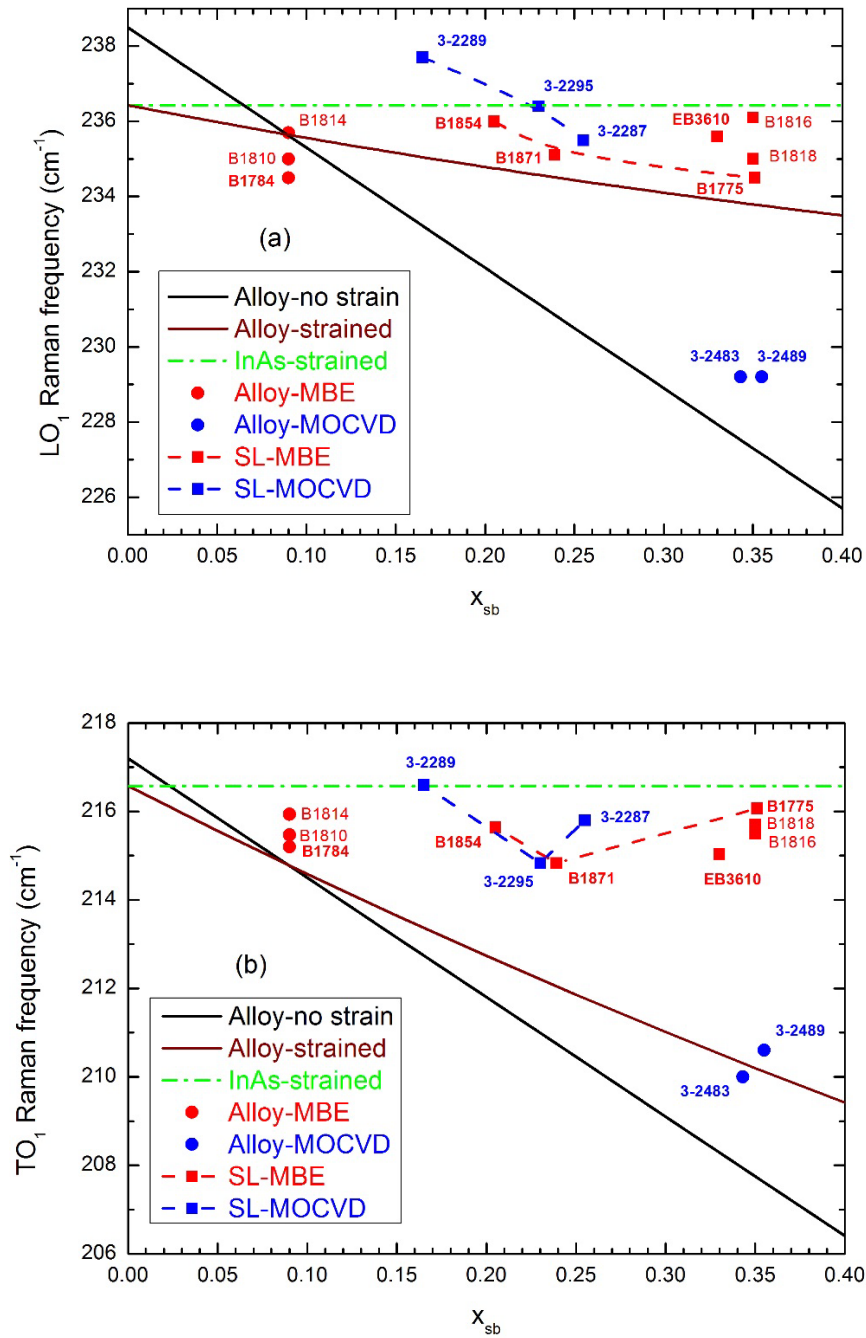

**Fig. 3**. Sb composition dependence of phonon frequency: (a) for the InAs-like LO phonon, (b) for the InAs-like TO phonon, calculated for alloys with and without strain (solid lines) and for strained InAs (dashed lines), and measured (symbols) for alloys (circular points) and superlattices (square points).

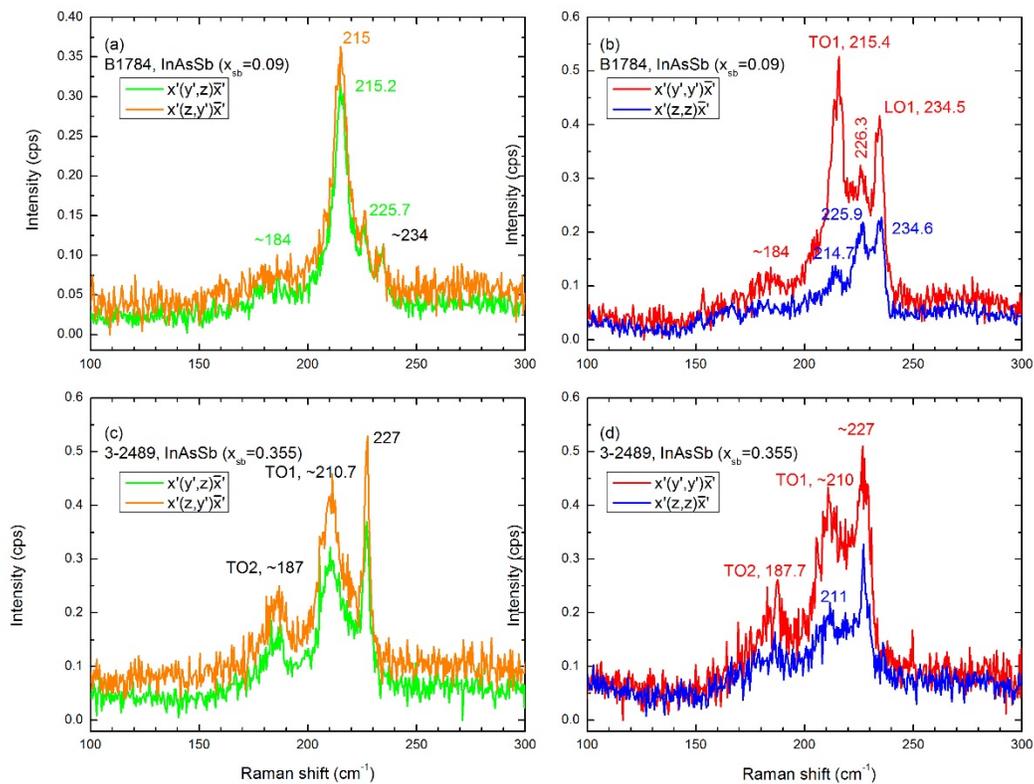

**Fig. 4.** (110) backscattering Raman spectra of InAs$_{1-x}$Sb$_x$ alloys in four polarization configurations for the same samples of Fig. 2. (a) B1784 (x$_{Sb}$ = 0.09) grown by MBE. (b) 3-2489 (x$_{Sb}$ = 0.355) grown by MOCVD.

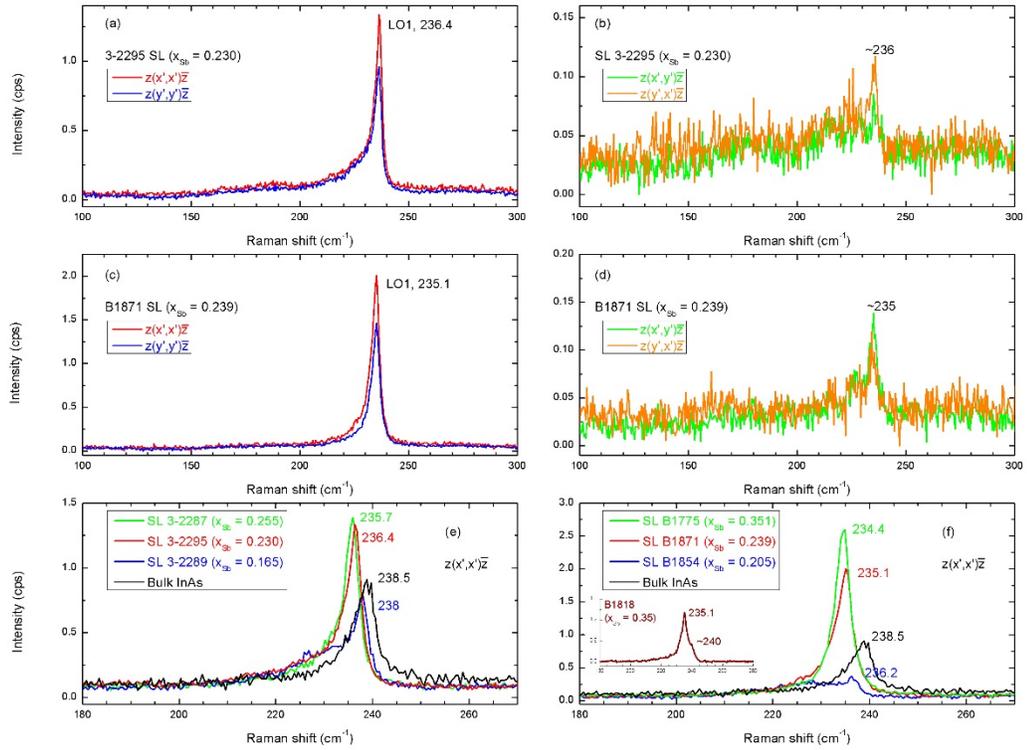

**Fig. 5.** (001) backscattering Raman spectra of InAs/InAs$_{1-x}$Sb$_x$ superlattices. (a) – (d) for two samples with similar structural parameters grown respectively by MOCVD and MBE, in four polarization configurations. (e) and (f), respectively, compare different superlattice samples grown by MOCVD and MBE with bulk InAs. The inset of (f) shows an additional MBE sample with Ga doping.

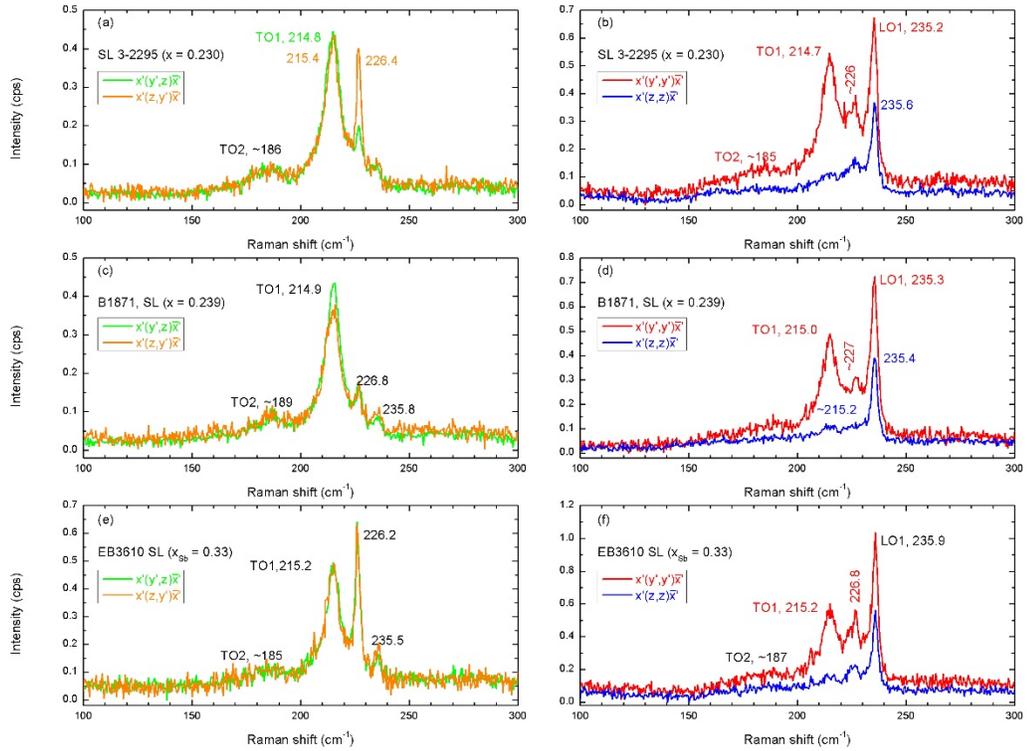

**Fig. 6.** (110) backscattering Raman spectra of InAs/InAs$_{1-x}$Sb$_x$ superlattices. (a) – (d) for two samples with similar structural parameters grown respectively by MOCVD and MBE, in four polarization configurations. (e) and (f) for another MBE sample grown by a different growth system.

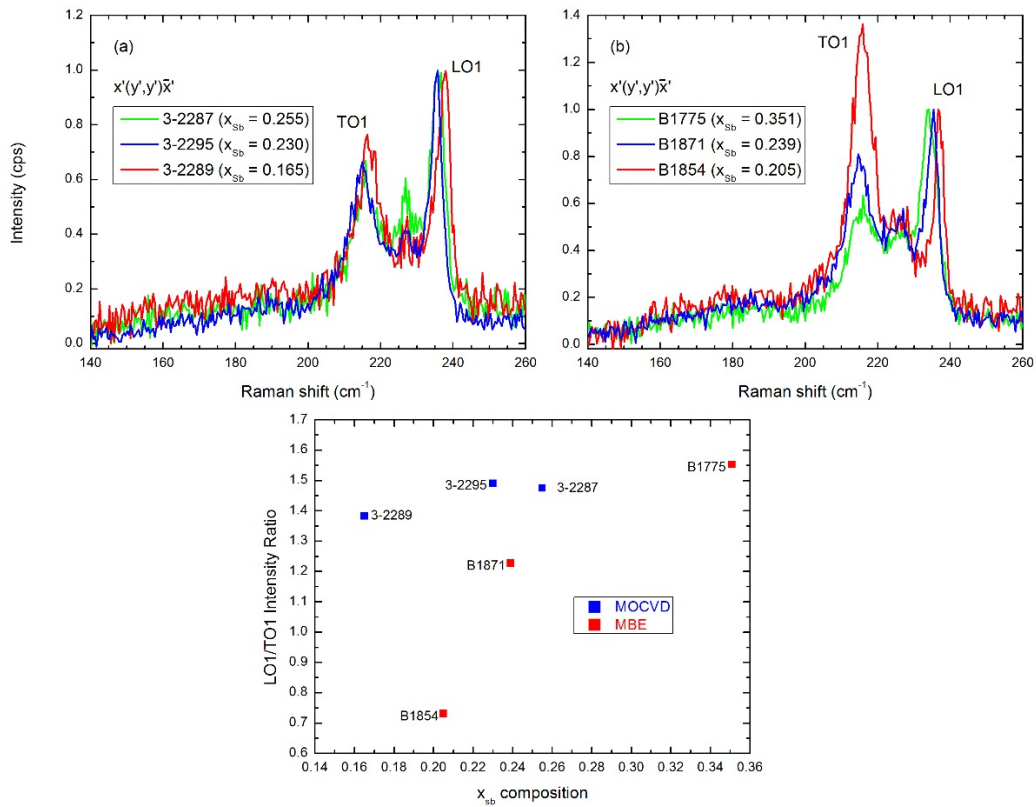

**Fig. 7** Sb composition dependence of the "forbidden" LO-like mode in InAs/InAs$_{1-x}$Sb$_x$ superlattices observed in the (110) backscattering geometry. The spectra are normalzed to the LO-like mode. (a) and (b), respectively, for MOCVD and MBE samples. (c) The peak intensity ratio of the LO-like mode to TO mode.

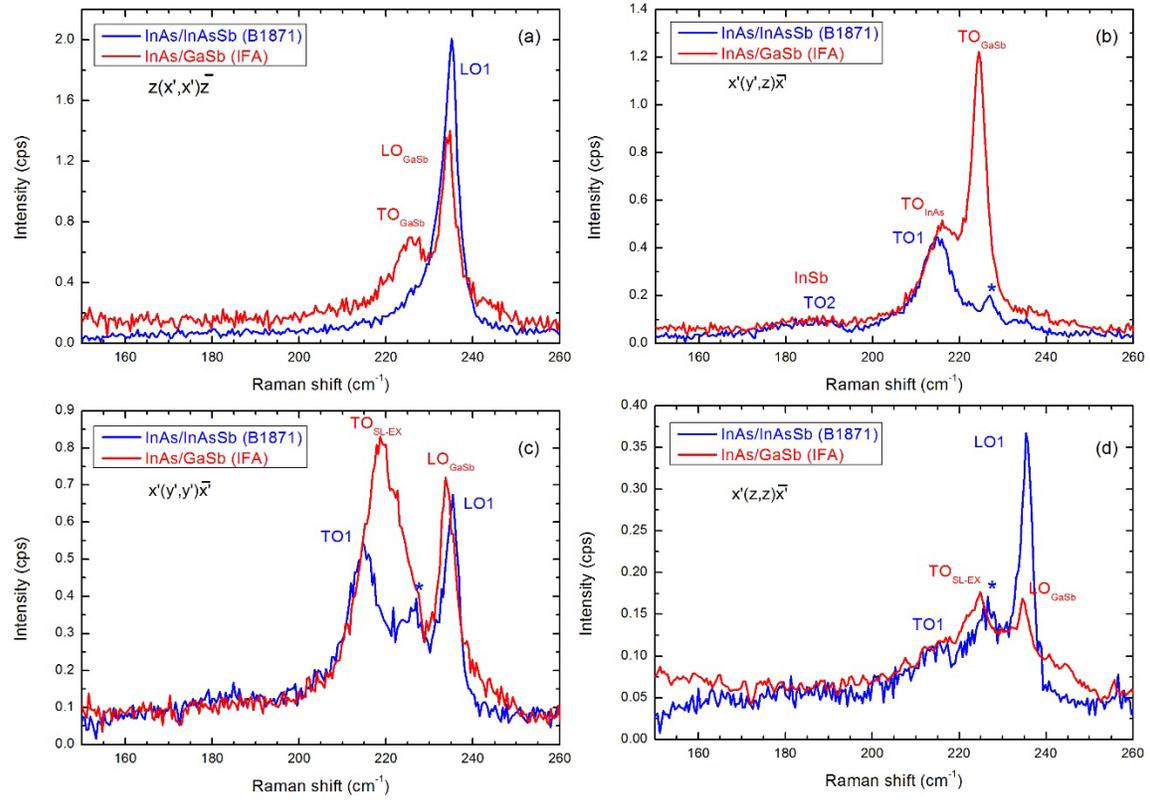

**Fig. 8** Comparison between InAs/InAs$_{1-x}$Sb$_x$ and InAs/GaSb superlattices: (a) for (001) backscattering, (b) – (d) for (110) backscattering in three polarization configurations. The peak indicated by "*" is from the GaSb substrate in the spectra for the InAs/InAs$_{1-x}$Sb$_x$ superlattice.